\documentclass[prl,aps,showpacs,twocolumn,preprintnumbers]{revtex4-1}
\usepackage{times,mathptmx,amssymb}
\usepackage{graphicx}
\usepackage[usenames,dvipsnames]{color}

\begin{document}
\title{Stochastic resonance on the transverse displacement of swimmers in an oscillatory shear flow}
\author{Francisca Guzm\'an-Lastra and Rodrigo Soto}
\affiliation{Departamento de F\'\i sica, FCFM, Universidad de Chile, Santiago, Chile}
\date{\today}


\begin{abstract}
Self-propelled microorganisms, such as unicellular algae or bacteria, swim along their director relative to the fluid velocity. Under a steady shear flow the director rotates in close orbit, a periodic structure that is preserved under an oscillatory shear flow. If the shear flow is subjected to small fluctuations produced by small irregularities in the microchannel  or by other swimmers nearby, the director dynamics becomes stochastic. Numerical integration of the swimmer motion shows that there is stochastic resonance: The displacement in the vorticity direction is maximized for a finite noise intensity. 
This transverse displacement resonance is observed when the displacement is coarse grained over several periods, although the director is preferentially oriented along the flow.
The resonant noise intensity is proportional to the oscillation frequency and independent of the shear rate.
The enhanced displacement can have effects on the transverse diffusion of swimmers and the rheology of the suspension.
\end{abstract}
\maketitle

\paragraph{Introduction.}

In recent decades, interest in the dynamics of self-propelled organisms has increased enormously. The interest is twofold, from continuum mechanics to describe the motion of single swimmers and from statistical physics to deduce collective behaviors that emerge due to their mutual interactions. 
Self-propelled organisms belong to what is called active matter, in which there is a continuous energy flux from some reservoir to produce motion; this energy is finally dissipated via viscosity or other similar means. From a statistical mechanics point of view, this energy flux puts active matter in out-of-equilibrium conditions.

Bacteria and unicellular algae are a particular kind of self-propelled organisms. Considering their micrometer scale and typical propulsion velocities, they are in the low-Reynolds-number regime, in which inertial effects are negligible in comparison with viscous ones.
In this category, the bacterium \emph{Escherichia coli} (\emph{E.~coli}) has been intensively studied, and much is known about its genetics, biological processes, and motility~\cite{EcoliMotion,RandomWalk}. 
At low Reynolds number, a swimmer can be modeled as a force dipole. Depending on wether the dipole is tensile or contractile, swimmers are classified as pushers or pullers, respectively. The distribution and orientation of the force dipoles in the fluid  have rheological effects. Indeed, elongated swimmers placed in a shear flow orient preferentially along the extensional direction. As a result, pushers, like the {\em E.coli}, reduce the fluid viscosity while pullers increase it~\cite{Hatwalne2004,Saintillan2010, Aranson,Aranson1,Saintillan2010b}. 

The motion of swimmers creates agitation of the fluid. This agitation can be visualized by placing solid particles as tracers. 
The induced tracer motion shows anomalous diffusion and in the long-time limit the induced motion is diffusive~\cite{Wu2000, Valeriani}. The induced diffusion has also been observed close to solid surfaces~\cite{Mino 2010}.
The fluid agitation affects the motion of nearby swimmers as well. When several swimmers interact, this agitation acts as a self-induced noise \cite{Aranson,Saintillan2008,Koch,Evans}.
In this Brief Report we study the effect of flow noise (either  self-induced noise or that from other sources) on the swimmer motion under an oscillatory shear. It will be shown that the noise enters multiplicatively in the swimmer equation of motion which is itself nonlinear. 

An interesting phenomenon called stochastic resonance (SR), in which  some response function is maximized for a finite noise intensity, can appear when a system is forced periodically. It first appeared that bistability, periodic forcing and random forces were necessary for the onset of SR~\cite{Hangii}. However, it later became clear that SR may appear in a large variety of systems, including linear systems subjected to multiplicative noise rather than to additive noise~\cite{Seki,Gitterman}. 
In this Brief Repport we present a SR in swimmers:  In an oscillatory shear flow, the displacement in the vorticity direction is maximized for a given noise intensity. 

\paragraph{Deterministic swimmer dynamics.}
Consider a self-propelled swimmer moving at low Reynolds number in a fluid. The swimmer propels with respect to the fluid at a velocity $V_0$ pointed by the director $\hat{n}$. If the swimmer's body is much smaller than the typical distance in which the fluid velocity changes, the Fax\'en correction can be neglected and the total velocity of the swimmer is
$\vec V=V_0\hat n+ \vec v$,
where $\vec v$ is the fluid velocity at the center of the swimmer~\cite{Happel 1965,KimKarilla}. The velocity gradients induce rotation of the swimmer, described by Jeffery's equation for the director vector~\cite{Jeffery}
\begin{equation}
\dot{\hat n}=(\mathbf I-\hat n\hat n)[\beta\mathbf E^{S} + \mathbf E^{AS}]\hat n. \label{jeffery1}
\end{equation}
Here $\mathbf E^{S}$ and $\mathbf E^{AS}$ are the symmetric and antisymmetric parts of the velocity gradient tensor $\mathbf E=\nabla \vec v$ and $\beta$ depends on the geometry of the swimmer. Limiting cases are $\beta=1$ for a rod-like body and $\beta=-1$ for a disk-like body. Here we are interested in the effects of imposing an oscillatory flow on the swimmer motion.

Under experimental conditions, it is possible to place bacteria in a Hele-Shaw geometry and impose an oscillatory flow. The resulting Poiseuille flow has a non uniform shear rate and, as a consequence of Eq.~(\ref{jeffery1}), the orientation dynamics of the bacteria depends on the vertical position in the cell. 
It is known that pushers tend to approach solid surfaces and swim close to them, remaining trapped by the surface for long periods~\cite{Ramia1993,Frymier1995,Lauga2006,Berke2008,Li2008,Dunstan2012}.  This fact  allows us to simplify the analysis. Close to the surface the shear rate is roughly constant and the same in the top and bottom surfaces. Therefore, a large majority of the bacteria swim in the flow with the same shear rate. To separate the effects of walls from the effect of the oscillatory shear flow, we will consider a simple shear flow throughout the fluid.
Although the equations of motion do not distinguish between pusher or puller swimmers, the previous discussion suggests that the simple shear approximation is of more relevance for pushers. 
Finally,  we  neglect the circular motion, with radius of some tens of microns, that flagellated swimmers develop near surfaces~\cite{Frymier1995, Lauga2006, Dunstan2012}.

\begin{figure}[htb]
\includegraphics[width=0.9\columnwidth]{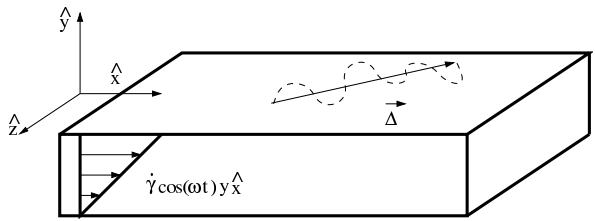}
\caption{Model setup. The swimmer moves in a fluid with an imposed oscillatory shear flow $\vec v^{\infty} = \dot{\gamma}^{\infty}\cos\left(\omega t \right)y\hat x$. The dashed curve depicts a possible trajectory and  the coarse grained displacement $\vec\Delta$ is evaluated for lapse times of several periods. }
\label{fig.setup}
\end{figure}

The swimmer is placed in an imposed simple oscillatory shear flow $\vec v^{\infty} = \dot{\gamma}^{\infty}\cos\left(\omega t \right)y\hat x$ (Fig. \ref{fig.setup}). Directions are such that $x$ is the flow direction, $y$ is the gradient direction, and $z$ is the vorticity direction. 

\begin{figure}[htb]
\includegraphics[width=.9\columnwidth]{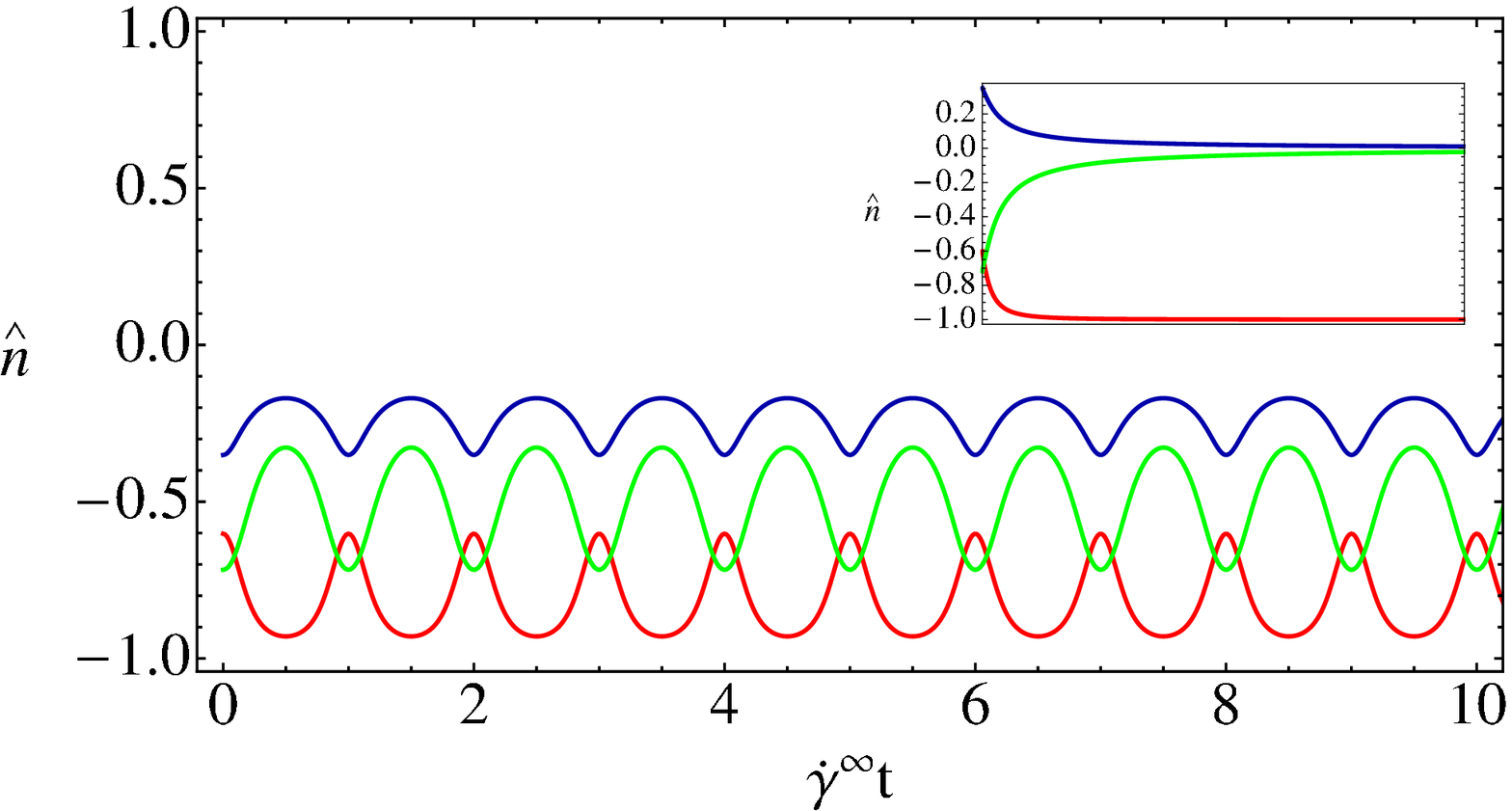}
\includegraphics[width=.9\columnwidth]{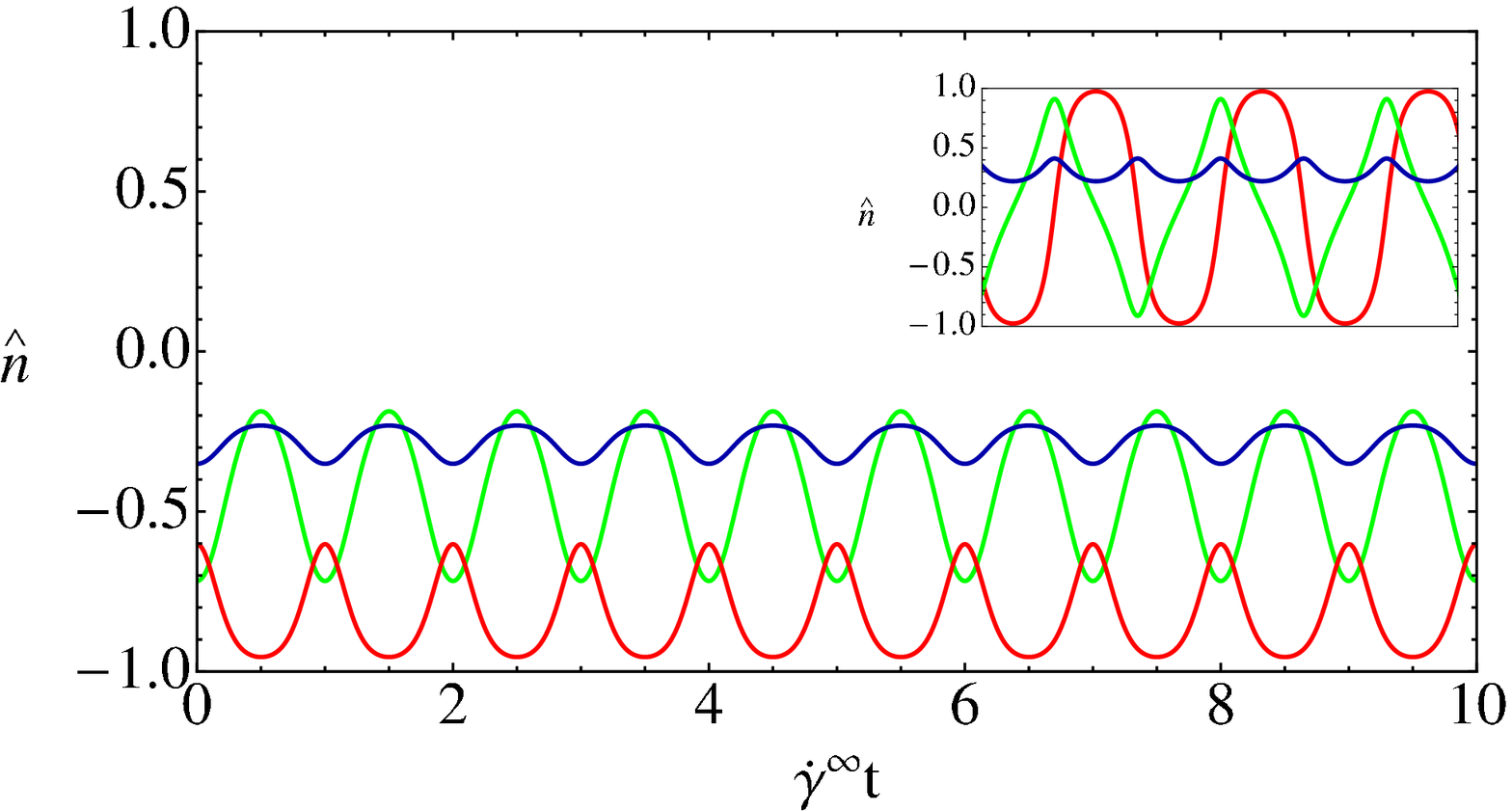}
\caption{(Color online). 
Time evolution of the components of the swimmer director: $n_x$ (gray lines, red online), $n_y$ (light gray lines, green online) and $n_z$ (dark gray lines, blue online). The swimmer geometric parameters are $\beta=1$ (top) and $\beta=0.6$ (bottom).
The main figures present the case of an imposed oscillatory shear flow with $\omega/\dot{\gamma}^{\infty}=2\pi$,  and the insets present the case of a steady shear flow with $\omega=0$. 
The curves depend on the initial conditions but are qualitatively similar for other initial conditions.}
\label{fig.orbits}
\end{figure}

In the case of a stationary flow ($\omega=0$), the temporal evolution of the director depends strongly on the value of $\beta$. Figure~\ref{fig.orbits} shows the  evolution of the director in the cases $\beta=0. 6$ and $\beta=1$. In the first case, the orbits are closed with a period proportional to $1/\dot{\gamma}^{\infty}$, while in the second case the orbit is open and the director orients asymptotically to the $\pm x$ direction. On the other hand, when the flow is oscillatory the orbits are always closed (periodic) and behave qualitatively similarly for different values of $\beta$ as shown in Fig. \ref{fig.orbits}. Based on these results we will analyze the case  $\beta=1$ in what follows. Other cases were studied showing qualitatively similar results~\cite{supmat}.

\paragraph{Noisy oscillatory flow.}
When a microscopic swimmer is placed in a flow, it is subjected to fluctuations of different origin. First, there is the thermal (Brownian) force that produces random reorientations and thermal components on the velocity. The effects of these have been largely studied and result in an effective diffusive motion \cite{RandomWalk} and smoothing of the Jeffery orbits~\cite{Saintillan2010,Saintillan2010b}.

The fluctuations on the velocity field (and velocity gradient) in which the swimmer moves are another source of noise. Velocity fluctuations do not produce large effects except on an added diffusive motion of the swimmer. However, as it will be shown, fluctuations in the velocity gradient  lead to a preferential displacement of the swimmer in the vorticity direction, an effect that is magnified for a given noise intensity.

In experiments, the noise in the velocity gradient can have several sources, three of which  we mention here. First, the imposed oscillatory flow can deviate from a perfect sinusoidal in an uncontrolled way. Second, the micro-channel can have (sub) micrometric roughness inducing velocity fluctuations in the Lagrangian frame of the fluid. Finally, other swimmers in the vicinity of the studied object create currents that are superimposed to the oscillatory flow. The intensities of these fluctuations depend on  experimental conditions and swimmer concentration~\cite{Saintillan2008,Koch,Evans}.

In the presence of noise, the shear rate tensor that appears in Jeffery's equation (\ref{jeffery1}) is
$\mathbf E(t) = \mathbf E^{\infty}(t) +\mathbf E^{\text{noise}}(t)$,
where $\mathbf E^{\infty}(t)=\dot\gamma^{\infty} \cos(\omega t) \hat{x}\hat{y}$ corresponds to the imposed oscillatory shear flow and $\mathbf E^{\text{noise}}(t)$ takes into account the velocity fluctuations. It is modeled as a tensor of white noise components of intensity $\Gamma$, with the trace subtracted to model an incompressible flow. That is, an intermediate tensor is built with components $F_{ij}$ satisfying  $\langle F_{ij}(t) F_{kl}(t')\rangle =\Gamma\delta_{ik}\delta_{jl}\delta(t-t')$. Then, the tensor $\mathbf E^{\text{noise}}$ has components $E^{\text{noise}}_{ik} = F_{ik} - F_{jj} \delta_{ik}/3$, and the resulting correlations are $\langle E^{\text{noise}}_{ij}(t) E^{\text{noise}}_{kl}(t')\rangle =\Gamma\left(\delta_{ik}\delta_{jl}-\delta_{ij}\delta_{kl}/3\right)\delta(t-t')$.
In summary, the position $\vec r$ and director $\hat{n}$ evolve according to the following equations
\begin{eqnarray} 
\frac{d\vec{r}}{dt}&=&V_0\hat n+\dot{\gamma}^{\infty} \cos(\omega t)y\hat x,\label{eq.motion1}\\
\frac{d\hat n}{dt}&=&(\mathbf I -\hat n\hat n)\left[\dot{\gamma}^{\infty}\cos(\omega t)\left(\begin{array}{ccc}
0&1&0\\
0&0&0\\
0&0&0 \end{array}\right)+ \mathbf E^{\text{noise}}\right]\hat n \label{eq.motion2}.
\end{eqnarray}
The noise intensity $\Gamma$ has units of inverse of time. It should be compared either with $\dot{\gamma}^{\infty}$ or with $\omega$ to quantify if the noise is large or small.
Considering the shear rate $\dot{\gamma}^{\infty}$, the oscillation frequency $\omega$ and the noise intensity $\Gamma$, two dimensionless parameters can be varied. We chose to fix $\dot{\gamma}^{\infty}$ and vary $\Gamma$ and $\omega$. Four  frequencies are used,  $\omega_1/ \dot{\gamma}^{\infty}=\pi/15$, $\omega_2 /\dot{\gamma}^{\infty}=\pi/10 $, $\omega_3/ \dot{\gamma}^{\infty}=\pi/5$, and $\omega_4/ \dot{\gamma}^{\infty}=2\pi/3$, while $\Gamma/\omega$ is varied in a wide range. We recall that the explored values of $\omega/\dot{\gamma}$ are experimentally feasible. For example, in a microfluidic device of cross section $L_z\times L_y=300\times50 \mu$m$^2$, with an imposed flux  $Q=5$ nl/s, the studied frequencies  scan the range $\omega\sim 0.6-13$ Hz.

In the equation for the swimmer director the noise is multiplicative. This can lead to complex phenomena in contrast to the effects produced by the additive noise that represents the thermal noise~\cite{Hangii}. The purpose of the present study is to describe the effect of varying the noise intensity.

\begin{figure}[htb]
\includegraphics[width=.8\columnwidth]{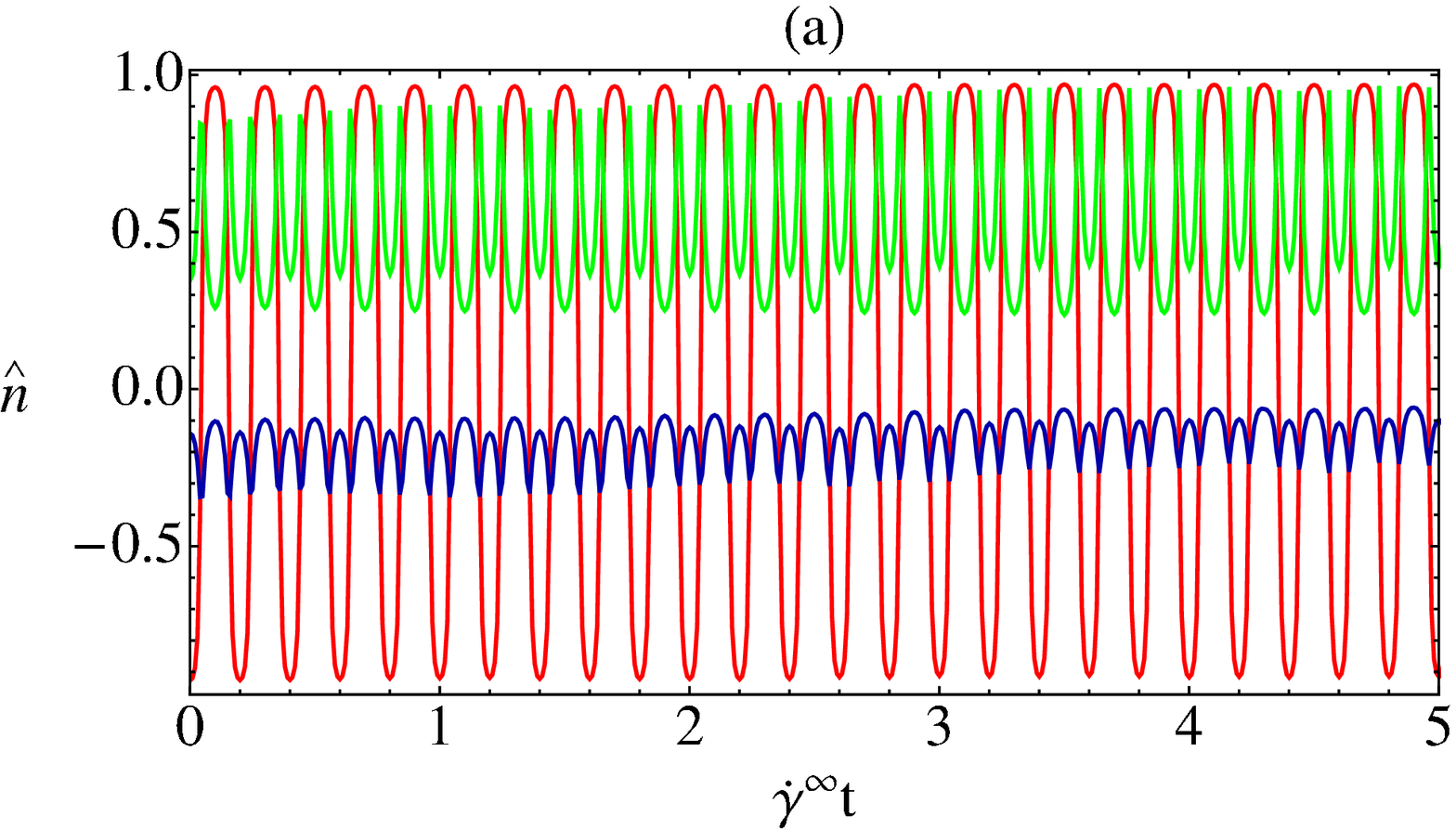}
\includegraphics[width=.8\columnwidth]{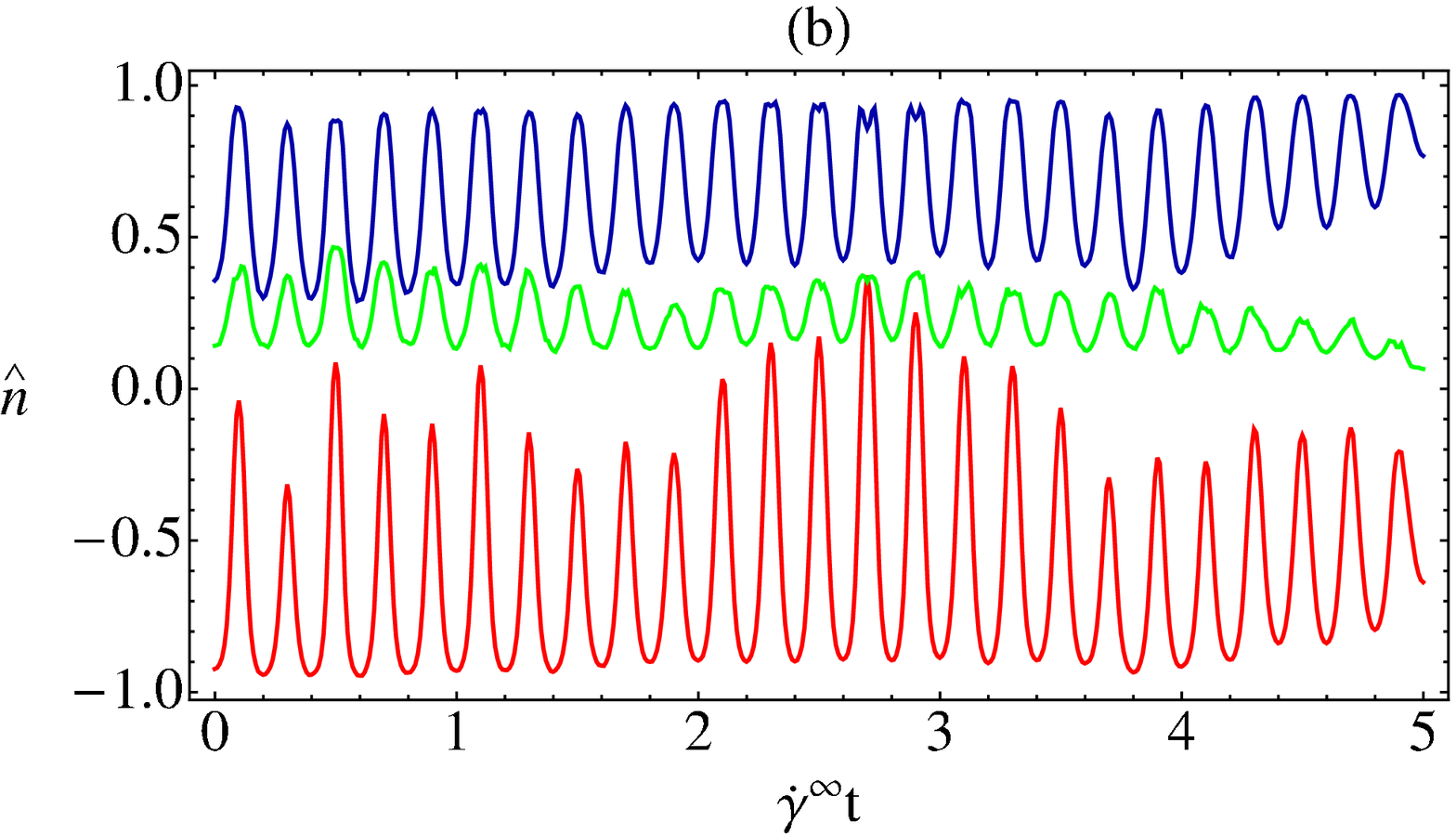}
\includegraphics[width=.8\columnwidth]{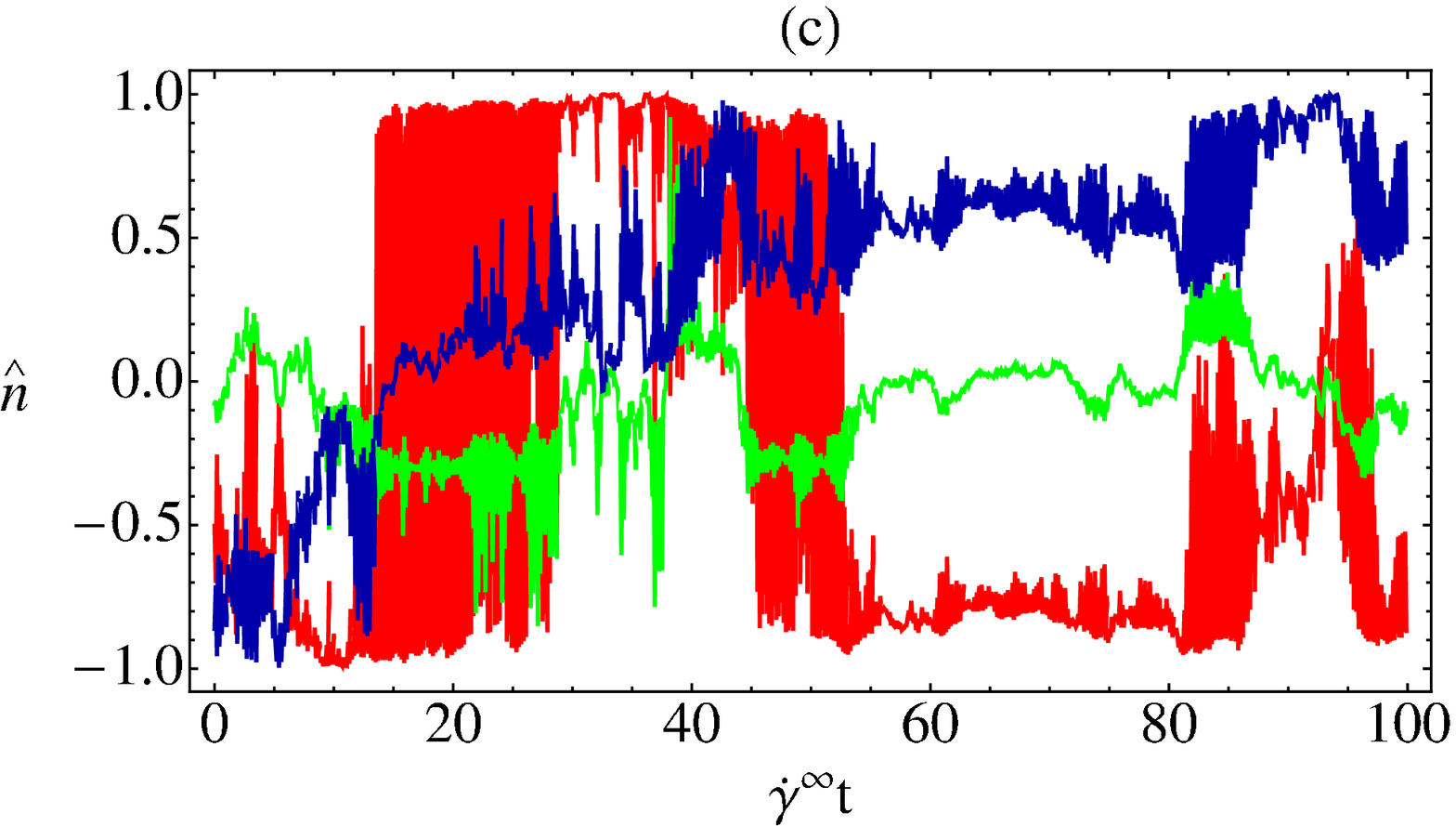}
\includegraphics[width=.8\columnwidth]{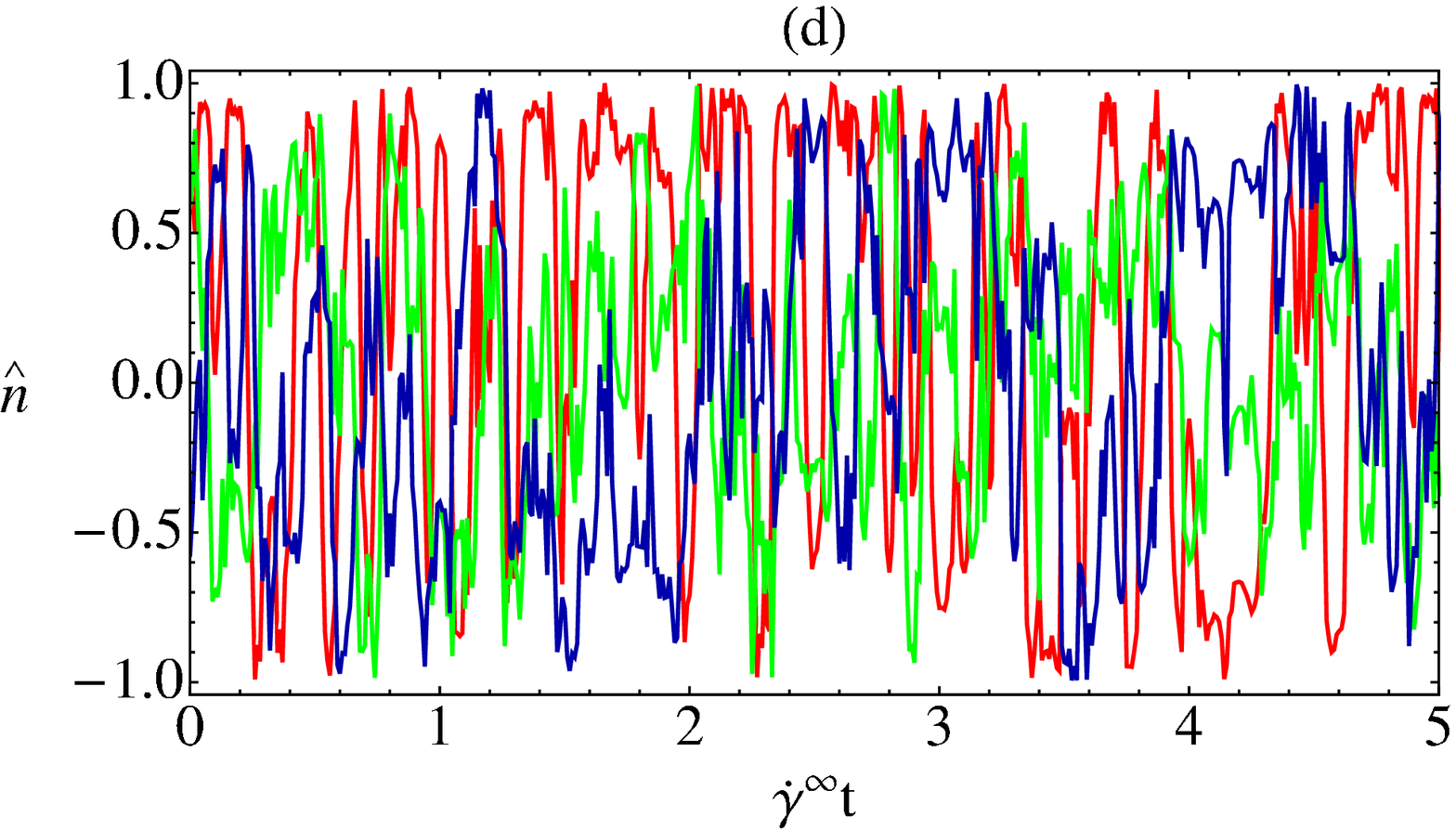}
\caption{
(Color online)
Time evolution of the components of the swimmer director under a noisy oscillatory shear flow: $n_x$ (gray lines, red online), $n_y$ (light gray lines, green online), and $n_z$ (dark gray lines, blue online). The noise intensities are (a) $\Gamma/\dot{\gamma}^{\infty}=0.001$, (b) $\Gamma/\dot{\gamma}^{\infty}=0.01$ (resonant noise intensity), (c) same value of $\Gamma$ for a longer time, and (d) $\Gamma/\dot{\gamma}^{\infty}=0.316$.  
The swimmer geometric parameter is $\beta=1$ and the frequency of the oscillatory shear flow is $\omega_2$. The curves depend on the initial conditions but are qualitatively similar for other initial conditions.}
\label{fig.orbitswnoise}
\end{figure}

 Equqations~(\ref{eq.motion1}) and (\ref{eq.motion2}) are interpreted according to the Stratonovich calculus, and they are numerically integrated using  Heun's predictor-corrector method~\cite{Platen,Heun}.
Figure \ref{fig.orbitswnoise} shows the evolution of the components of the director vector for different noise intensities. For small noise intensities, the Jeffery orbits are slightly perturbed. For long times (not shown in the figure) the director oscillates around the same direction fixed by the initial conditions. 
For large noise intensities the periodic structure is completely lost and the director performs a random motion.
In the case of an intermediate noise intensity, the orbits preserve some periodicity, but it is lost after some periods. In the long-time term the swimmer oscillates for long periods around some orientations until it switches to a new orientation. The transitions take place at random times and the new orientations are also random.

To quantify the degree of orientation, averages of the quadratic components of the director vector are considered. The swimmer preferentially orients along the $x$ axis except  for large noise intensities ($\Gamma/\dot{\gamma}^{\infty}\sim 1$) when the swimmer orients isotropically (see Supplemental Material \cite{supmat}).  
Note that in the case of passive elongated fibers, it has been found that they preferentially orient along the $z$ axis \cite{Franceschini}. However, in that case, the alignment is produced by excluded volume effects that are absent in our case because we consider an isolated swimmer.

Although the swimmer orients principally on the $x$ axis (positive and negative directions), the mean displacement can be in a different direction as the oscillations can lead to cancellations in the $x$ direction. To subtract the effect of the rapid oscillations, the displacement vector $\vec\Delta$ is computed for lapse times of $30$ periods, and later divided by 30. This number of periods is sufficiently large  to obtain a coarse-grained description of the displacements, averaging over the back-and-forth motion induced by the Jeffery orbits and the shear flow oscillations, but small enough to capture the coherent motion shown in Fig.~\ref{fig.orbitswnoise}(c). From the displacement vector, averages $\langle \Delta_y^2\rangle$ and $\langle \Delta_z^2\rangle$ are computed. The average $\langle \Delta_x^2\rangle$ is not well defined because it depends on the streamline in which the swimmer is located.

\begin{figure}[htb]
\includegraphics[width=.9\columnwidth]{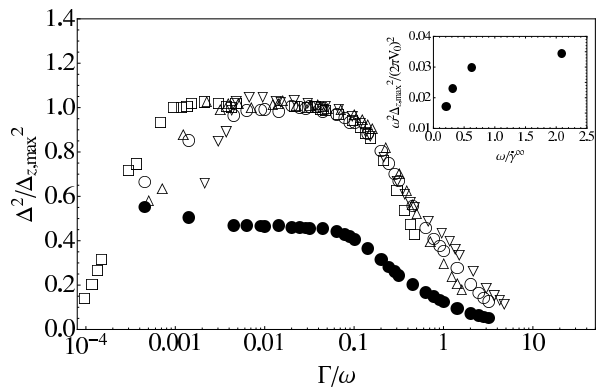}
\caption{
Average squared coarse grained displacements as a function of the noise intensity $\Gamma$. The linear vertical scale has been rescaled for each frequency to the maximum value of $\langle\Delta_z^2\rangle$. The oscillation  frequencies are $\omega_1$ (triangles), $\omega_2$ (circles),  $\omega_3$ (squares), and  $\omega_4$ (diamonds). $\langle\Delta_z^2\rangle$ in open symbols and $\langle\Delta_y^2\rangle$ in solid symbols (only one frequency is shown, others are similar).
Inset:  Maximum squared coarse-grained displacement in the $z$ direction as a function of the oscillation frequency, normalized with the distance traveled in one period.
}
\label{fig.DeltaZ2}
\end{figure}

The average squared displacements are shown in Fig. \ref{fig.DeltaZ2}  as a function of the noise intensity. The displacement in the vorticity direction--$\langle \Delta_z^2\rangle$--shows a maximum for a small but finite noise intensity. The swimmers show an enhancement of the transverse motion even though the director is mainly oriented along the $x$ direction. Figure~\ref{fig.DeltaZ2} shows the resonant displacement normalized with the distance traveled in one period, presenting a  weak increase with the oscillation frequency. The squared displacements in the $y$ direction are smaller than those in the $z$ direction and show no maximum.

The observed phenomenon is a stochastic resonance in which the response (transverse displacement) is maximized for a finite noise intensity~\cite{Hangii,NoisyOscillator,Gitterman,Seki}. Smaller noises lead to slightly perturbed Jefferry's orbits and large noises produce a complete isotropic response. At the resonant noise intensity the trajectory has the appearance shown in Fig.~\ref{fig.orbitswnoise}(c). It should be mentioned that the resonant noise intensity is small, $\Gamma^{\rm res}\ll\dot{\gamma}^{\infty},\omega$. This is  responsible for the large transition times observed in Fig.~\ref{fig.orbitswnoise}(c).
The resonance curves for different frequencies collapse when plotted against $\Gamma/\omega$ and rescaled to their maximum value. 
The resonance is wide and rather flat, making it difficult to identify the resonance noise intensity $\Gamma^{\rm res}$ with precision. It lays in the range $\Gamma^{\rm res}=(8-30)\times 10^{-2}\omega$.  

Stochastic resonance was observed in linear systems (i.e., without bistability), subjected to multiplicative noise as long as the noise had some finite correlation time~\cite{Seki,Gitterman}. In the present case,  the noise correlation time is zero but two aspects could have allowed to overcome this limitation. First, the equations are non-linear and no simple analysis excludes SR in this case. Secondly, this is a coupled system of equations and it is known that in this case finite correlation times can develop as in the Langevin modeling of the Ornstein-Uhlenbeck noise~\cite{vanKampen}.

\paragraph{Perspectives.} 
The dynamics of  a single swimmer moving at low Reynolds number is studied in the presence of an oscillatory shear flow. In microfluidic devices,  it is known that bacteria approach and swim close to the solid surfaces. Therefore, the flow acting on a swimmer can be  modeled as a simple shear. When flow fluctuations are taken into account, the resulting equation for the director vector is non-linear and the noise is multiplicative. When analyzing the coarse-grained displacement, stochastic resonance is observed. The displacement in the vorticity direction, transverse to the flow, is maximized for a finite noise intensity. The resonance is wide, therefore it is difficult to obtain precisely the resonant noise intensity $\Gamma^{\rm res}$. Based on the results for four different frequencies it is found that $\Gamma^{\rm res} \propto\omega$, with a small proportionality constant.

The observed stochastic resonance implies that if the transverse motion were diffusive, the transverse diffusivity and mixing could be maximized by varying the noise intensity. Also, the preferential orientation of the director along the flow can have rheological effects on the frequency-dependent viscosities, an analysis that is being performed and will be published elsewhere.

In the model, we have not considered the rapid reorientation of bacteria (tumbling). The effect of tumbling would be to render the motion more isotropic, thus reducing the resonance amplitude. In addition, the hydrodynamic interactions with  surfaces could affect the resonance.

Finally, a more detailed analysis of the stochastic process is necessary to identify the key ingredients that produce the resonance. It is, however, complex, as there are several degrees of freedom and two intrinsic frequencies, $\omega$ and $\dot{\gamma}$. A simpler toy model could provide insight.

\paragraph{Acknowledgments.}
We thank M.G. Clerc, E. Altshuler, E. Clement and G. Mi\~no for valuable discussions.  This research is supported by Fondecyt 1100100, Anillo ACT 127, and ECOS C07E07 grants. F.G. acknowledges the support of a CONICYT grant.

\end{document}